\documentclass[twocolumn]{aastex62}

\usepackage{graphicx} 
\usepackage{array, geometry}
\usepackage{threeparttablex, booktabs}
\usepackage{natbib}
\usepackage{amsmath}
\usepackage{xcolor}

\submitjournal{PASP}
\shorttitle{Multipole analysis}
\shortauthors{Leahy, Ranasinghe, Hansen, Filipovic \& Smeaton}

\begin{document}

\title{Multipole Analysis and Application to Supernova Remnant X-ray Images}

\correspondingauthor{Denis Leahy}
\email{leahy@ucalgary.ca}

\author{D.A. Leahy}
\affiliation{Department of Physics and Astronomy, University of Calgary, Calgary, AB T2N 1N4, Canada}

\author{S. Ranasinghe}
\affiliation{Department of Physics and Astronomy, University of Calgary, Calgary, AB T2N 1N4, Canada}

\author{J. Hansen}
\affiliation{Department of Physics and Astronomy, University of Calgary, Calgary, AB T2N 1N4, Canada}

\author{M. D. Filipovi\'c}
\affiliation{Western Sydney University, Locked Bag 1797, Penrith South DC, NSW 2751, Australia}

\author{Z. Smeaton}
\affiliation{Western Sydney University, Locked Bag 1797, Penrith South DC, NSW 2751, Australia}

\begin{abstract}

We develop a multipole analysis method for images with a circular boundary, then apply it to supernova remnant (SNR) images. The morphology of SNR images is related to several factors, including the inhomogeneities of the supernova ejecta and of the circumstellar medium in which the ejecta and shock wave travel. 
The current multipole method corrects some errors in a previously presented method, and applies the new analysis to test for differences in X-ray image morphology between Type Ia and core-collapse type SNRs.
We find there is no clear difference between the two SNR types in morphology as measured by multipole moments.
\end{abstract}


\section{Introduction}\label{sec:intro}

Supernova remnants (SNR) have a large impact on the interstellar medium in galaxies, and can shape the overall structure of the ISM (e.g., see the review paper \citealt{2005ARA&A..43..337C}). 
The evolution of a SNR up to the time it becomes radiative is reviewed by \citet{1999ApJS..120..299T}, and after it becomes radiative by \citet{1988ApJ...334..252C}.
These assume spherical symmetry, and a code to calculate the evolution of a spherical during all phases, including the early self-similar ejecta dominated phase \citep{{1982ApJ...258..790C}} was presented by \citet{2019AJ....158..149L}. 

The evolution of a SNR in non-uniform media has been modelled only in a few studies, e.g. in a stellar wind bubble (still spherically symmetric) and in SNR which encounters a linear jump in density  \citep{1996ApJ...471..279D,2008A&A...478...17F}. 
Interior instabilities that develop during evolution of an SNR in the wind blown bubble of its massive progenitor star have been simulated by \cite{2007ApJ...667..226D}.
The comparison of non-uniform or non-spherical simulations with observations is difficult, in part because of the large possible number of ways to set the initial conditions.

From a phenomenological point of view, an analysis of an image of a SNR can inform the degree of non-sphericity of a SNR.
The SNR is optically thin, including for radio and X-ray wavelengths.
The image is a 2 dimensional (2D) projection of the 3 dimensional volume of the SNR and one measures the non-uniformity of the 2D projection.
Most often the non-uniformity has been described only qualitatively, however \cite{2009ApJ...706L.106L} introduced a quantitative measure of the non-uniformity of a 2D image.
\textit{Chandra} X-ray images of 17 SNRs, of which 7 were Type Ia and 10 were core-collapse type, were analyzed. 
The method utilized what was called the power-ratio method or PRM \citep{1995ApJ...452..522B}.
The PRM is a series expansion for the 2-dimensional projected gravitational potential $\Psi(\rho,\phi)$ using 2D polar coordinates ($\rho,\phi$). 
This assumes that $\Psi(\rho,\phi)$ satisfies the equation:
\begin{equation}
\nabla^2 \Psi(\rho,\phi)= 4\pi G \Sigma(\rho,\phi)
\end{equation}
The series expansion for $\Psi(\rho,\phi)$ was given by Equation 2 of \cite{1995ApJ...452..522B}.
\cite{2009ApJ...706L.106L} found a clear difference between Type Ia and core collapse SNRs, with core collapse SNRs clearly separated in the $P_2/P_0$ vs. $P_3/P_0$ plane, where $P_0$, $P_2$ and $P_3$ are the multipole expansion powers of order 0, 2 and 3, respectively.

A more recent study using the PRM was given by \cite{2011ApJ...732..114L}, who expanded the sample for PRM analysis of \textit{Chandra} images of SNRs to 10 Type Ia, 13 core-collapse type and one other (SNR G344.7-0.1), and verified the results of \cite{2009ApJ...706L.106L}. 
They included a separate wavelet transform analysis of the X-ray line images of the SNRs to measure the distribution of emission line size scales, with the result that there was no clear difference between the different SNRs (except for W49B, which was attributed to a jet-driven explosion mechanism).  
\cite{2019arXiv190911803R} applied the PRM analysis to radio images of SNRs and found no significant difference between Type Ia and core collapse SNRs, in contrast to what was found earlier for X-ray images. 

In this paper we re-examine the use of the PRM for analysis of SNR images. 
Section 2 goes over the underlying mathematics of the multipole expansion for images and shows the PRM is not applicable to expansion of images in general, mainly because it does not use a complete set of basis functions, nor are they orthogonal.
Section 3 introduces the sample of X-ray images of SNRs. 
The multipole analysis for SNR images is carried out in Section 4. 
In Section 5, we discuss the results, which show no clear difference in multipoles between Type Ia and core-collapse SNRs. 
The conclusions are summarized in Section 6.

\section{Methods}\label{sec:meth}

\subsection{Complete Sets of Orthogonal Multipoles}\label{complete}

Complete sets of solutions of self-adjoint differential equations form complete sets of functions \citep{mathphys2,mathphys1}. 
The Gramm-Schmidth ortho-normalization procedure can be applied to any complete set of functions to turn them into a complete set of orthogonal and normalized functions.

To completely represent an image (a 2D function) one requires a complete set of 2D functions with the same boundaries as the image. 
For example, a rectangular image can be represented in terms of Fourier series in rectangular (x,y) coordinates.
However, many astronomical images have a basic circular symmetry, with small or large perturbations, so are better represented in 2D plane polar coordinates with a circular boundary.
For the latter case, the first requirement is to find a complete set of orthogonal functions in 2D polar coordinates $0<\rho<R$, $0<\phi<2\pi$, where $R$ is the radial coordinate of the boundary. 
By rescaling the outer boundary to unity, one has $r=\rho/R$, $0<r<1$.

An appropriate self-adjoint differential operator in plane polar coordinates is $\nabla^2 \Phi(r,\phi)=0$. 
The boundary conditions are $\Phi(r,\phi)=\Phi(r,\phi+2\pi)$ (i.e. single valued); and $\Phi(0,\phi)$ and $\Phi(1,\phi)$ are finite. 
A complete set of functions consists of products of radial functions $r^n$, $n=0,1,2...\infty$, and angular functions. 
The angular functions are $exp(\pm im\phi)$, $m=0,1,2...\infty$ or equivalently $cos(m\phi)$, $m=0,1,2...\infty$ and $sin(m\phi)$, $m=1,2...\infty$. 
Here we choose the $cos(m\phi)$ and $sin(m\phi)$ set instead of the $exp(\pm im\phi)$ set\footnote{The $cos(m\phi)$ and $sin(m\phi)$ set are orthogonal over the interval $0<\phi<2\pi$. They are orthonormal if multiplied by $1/\sqrt(2\pi)$ for $m=0$ or $1/\sqrt(\pi)$ for $m>0$.}. 
However, the radial functions $r^n$, $n=0,1,2...\infty$, are not orthogonal.
Gramm-Schmidt ortho-normalization applied to $r^n$ over the interval $0<r<1$ yields the Shifted Legendre polynomials, 
$\sqrt{2n+1}~ P_{n}^{*}(r)$.
Thus $\sqrt{2n+1}~P_{n}^{*}(r)$ are the appropriate radial basis functions.

The resulting complete and orthonormal basis functions are:
\begin{equation}
\label{eqn.2}
fc_{n,0}(r,\phi)=\sqrt{\frac{2n+1}{2\pi}}~ P_{n}^{*}(r)
\end{equation}
$n=0,1,2...\infty$, m=0; 
\begin{equation}
\label{eqn.3}
fc_{n,m}(r,\phi)=\sqrt{\frac{2n+1}{\pi}}~P_{n}^{*}(r) ~cos(m\phi)
\end{equation}
$n=0,1,2...\infty$, $m=1,2...\infty$; and 
\begin{equation}
\label{eqn.4}
fs_{n,m}(r,\phi)=\sqrt{\frac{2n+1}{\pi}}~P_{n}^{*}(r) ~sin(m\phi)
\end{equation}
$n=0,1,2...\infty, m=1,2...\infty$.
These are also known as 2D multipole functions.
$n$ is known as the radial order of the multipole, and $m$ as the angular order of the multipole.

\subsection{Application to Representation of Images}\label{sec:application}

Any image with circular boundary can be represented in terms of the above complete set of orthogonal multipole functions.
The standard multipole basis functions are those above: $fc_{n,m}(r,\phi)$ and $fs_{n,m}(r,\phi)$, and the multipole expansion of an image $(F(r,\phi))$ is written:
\begin{eqnarray}
\label{eqn.5}
F(r,\phi)&=&\sum_{n=0,m=0}^{\infty,\infty}a_{n,m}~fc_{n,m}(r,\phi)   \nonumber \\
& &+\sum_{n=0,m=1}^{\infty,\infty}a_{n,m}~fs_{n,m}(r,\phi)   \nonumber \\ 
\end{eqnarray} 
with the multipole coefficients given by:
\begin{equation}
\label{eqn.6}
ac_{n,m}=\int_{r=0}^{1} dr \int_{\phi=0}^{2\pi}~d\phi ~F(r,\phi) ~fc_{n,m}
\end{equation}
$n=0,1,2...\infty$, $m=0,1,2...\infty$
and \begin{equation}
\label{eqn.7}
as_{n,m}=\int_{r=0}^{1} dr \int_{\phi=0}^{2\pi}~d\phi ~F(r,\phi)~ fs_{n,m}
\end{equation}
$n=0,1,2...\infty$, $m=1,2...\infty$.
Because the images to be analyzed are real and the multipole functions are real, the coefficients of the multipole expansion of the image are real. 

If the set of functions are not orthogonal and complete, the relations Equations (5), (6) and (7) are not valid and one cannot represent the image in terms of the basis functions.
The PRM method basis functions are neither orthogonal nor complete so they cannot be used to correctly represent an image.

A summary of the multipole representation consists of the multipole power-ratios.
For radial order n and angular order m the power-ratio (often called the power) $P(n,m)$ is given by:
\begin{equation}
\label{eqn.8}
P_{n,m}=(ac_{n,m}^2+as_{n,m}^2)/a_{0,0}^2   
\end{equation}
where $as_{n,0}$ is defined as $0$ to avoid having a separate definition for the case $m=0$ for the sine basis functions. 
These are normalized to be independent of the total flux of the image, measured by the $a_{0,0}$ coefficient.
The powers are a measure of the brightness in an image in each radial and angular order. 
They do not include the phase information of orders, thus do not completely specify the image. 
One requires the full set of coefficients $ac_{n,m}$ and $as_{n,m}$ in order to represent the image using Equation~(5) above.

One can define radial multipole powers which are summed over all angular orders $m$:
 \begin{equation} \label{eqn.9}
P_{rad,n}=\sum_{m=0}^{\infty}(ac_{n,m}^2+as_{n,m}^2)/a_{0,0}^2   
\end{equation}
and angular powers which are summed over all radial orders $n$:
 \begin{equation} \label{eqn.10}
P_{ang,m}=\sum_{n=0}^{\infty}(ac_{n,m}^2+as_{n,m}^2)/a_{0,0}^2   
\end{equation} 
The advantage of defining radial powers and angular powers is that it allows one to discern whether the difference between two images is attributed to their radial structure (difference in $P_{rad,n}$) or attributed to their angular structure (difference in $P_{ang,m}$).

\begin{figure*}
    \centering
    \includegraphics[width=0.19\linewidth]{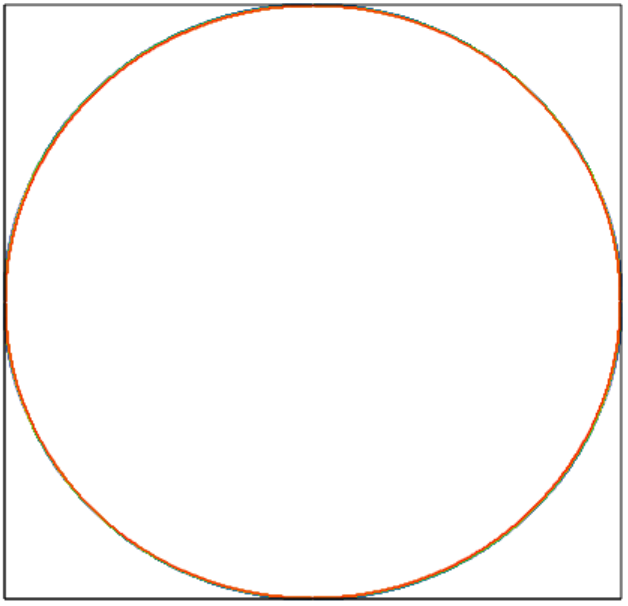} 
    \includegraphics[width=0.19\linewidth]{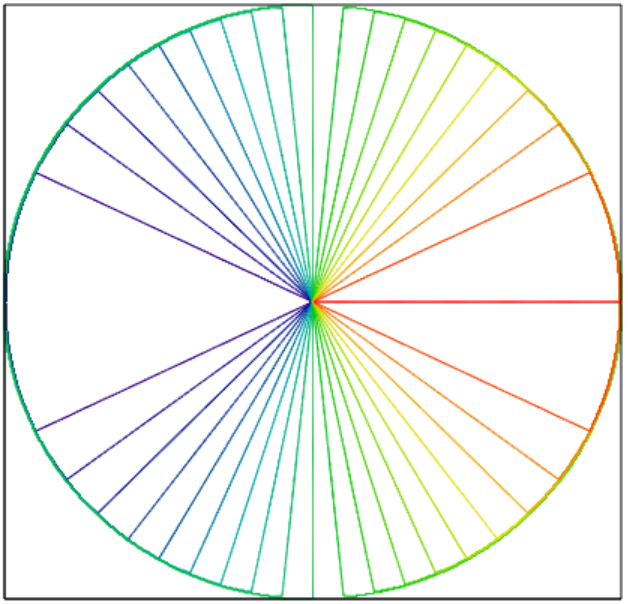} 
    \includegraphics[width=0.19\linewidth]{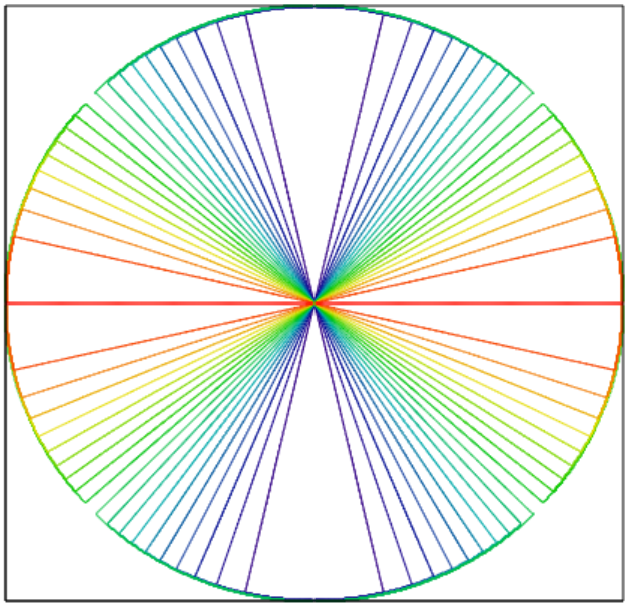} 
    \includegraphics[width=0.19\linewidth]{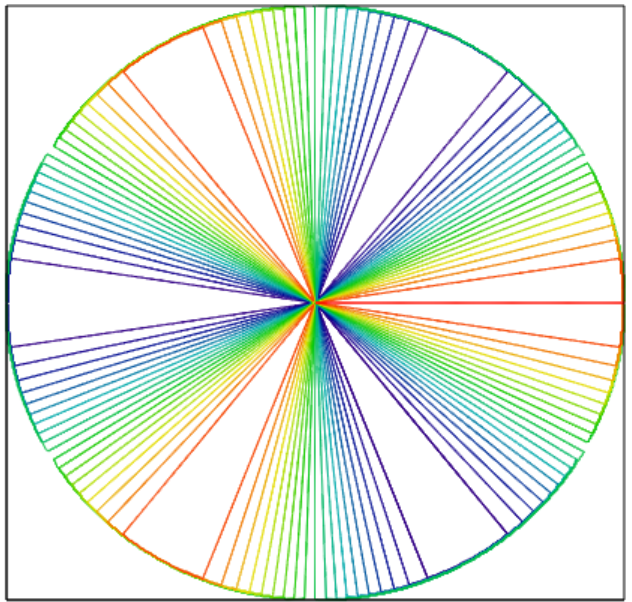} 
\includegraphics[width=0.19\linewidth]{P00C.png}
    \includegraphics[width=0.19\linewidth]{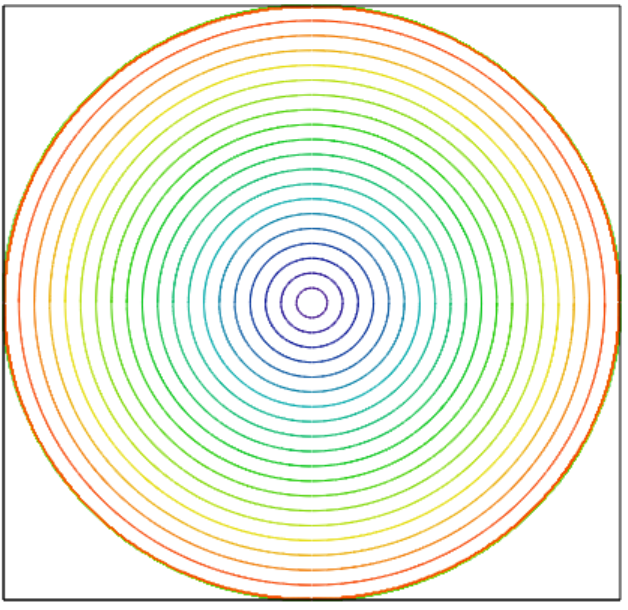}  
    \includegraphics[width=0.19\linewidth]{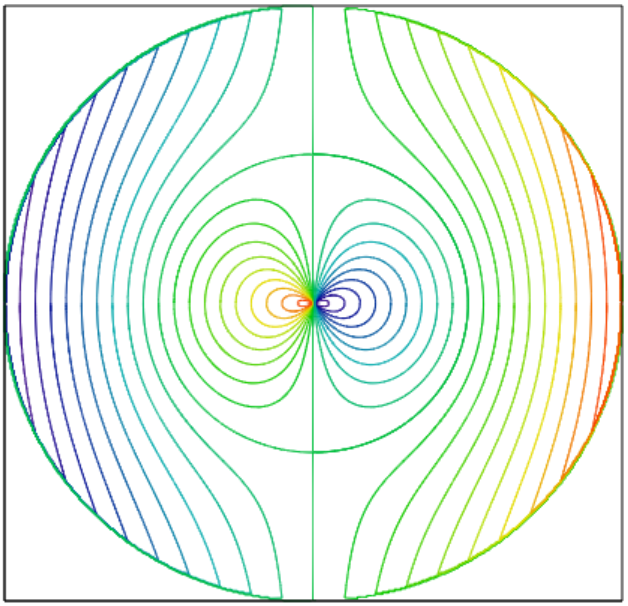}
    \includegraphics[width=0.19\linewidth]{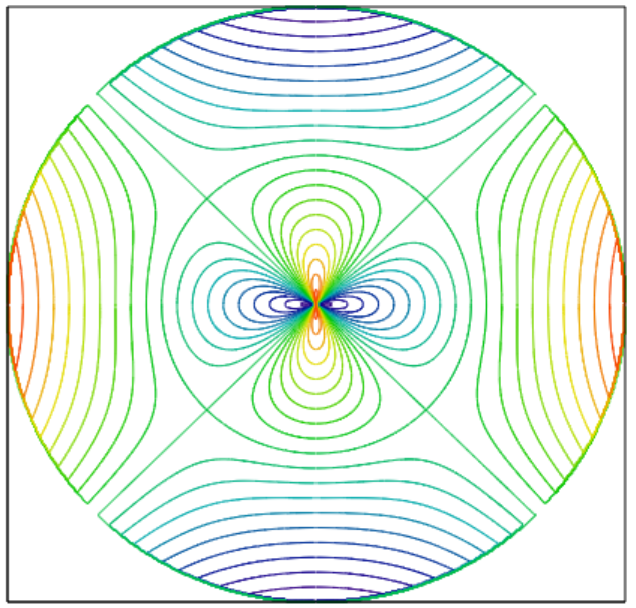}
    \includegraphics[width=0.19\linewidth]{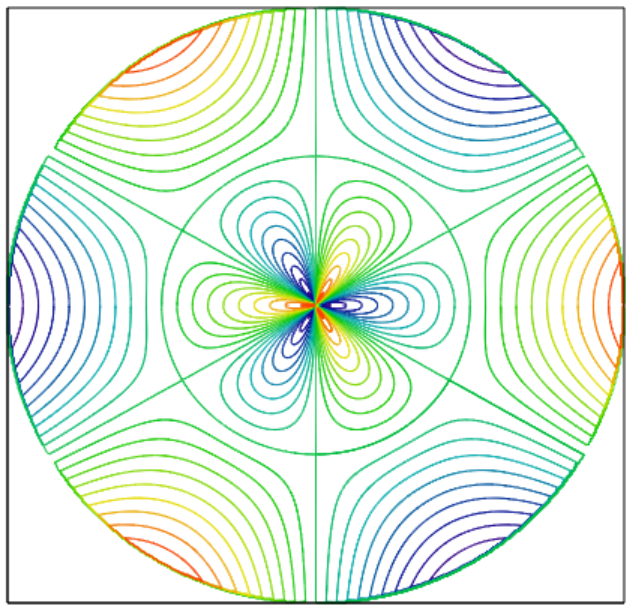} 
\includegraphics[width=0.19\linewidth]{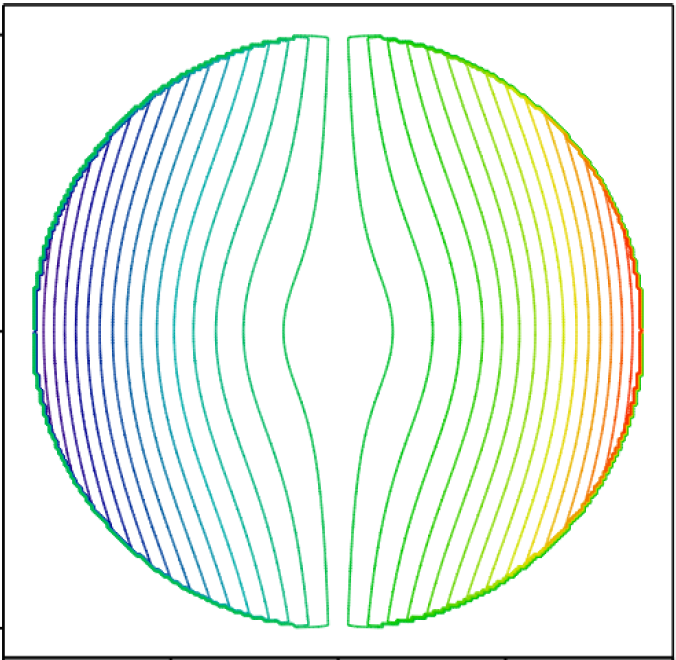}
    \includegraphics[width=0.19\linewidth]{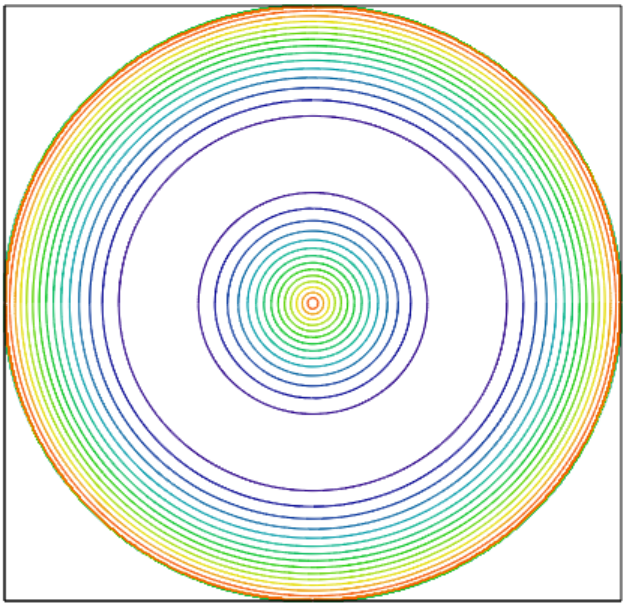}  
    \includegraphics[width=0.19\linewidth]{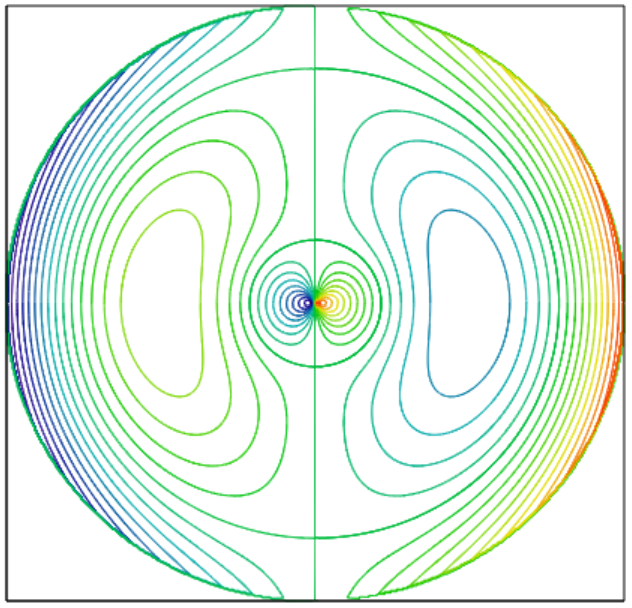}
    \includegraphics[width=0.19\linewidth]{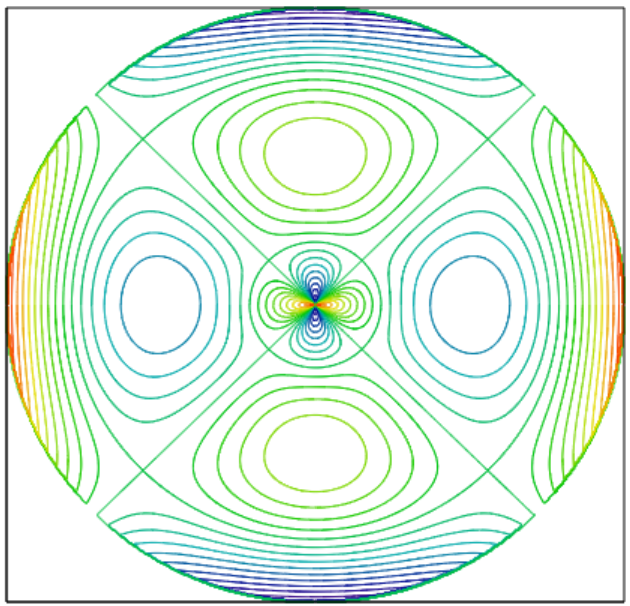}
    \includegraphics[width=0.19\linewidth]{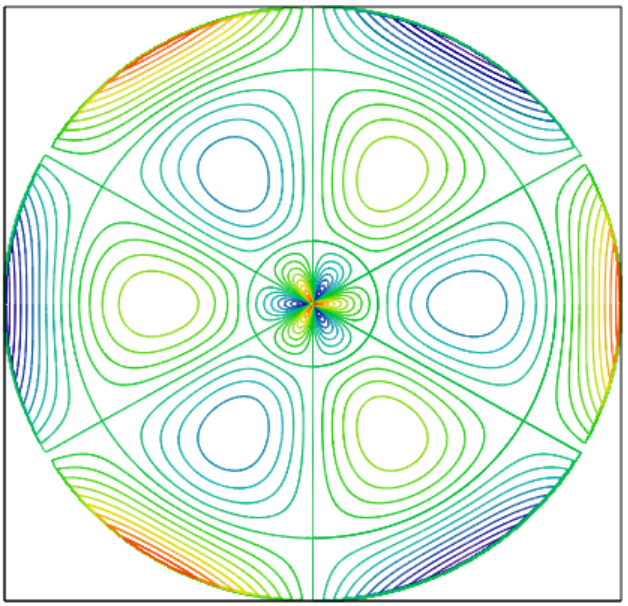}
\includegraphics[width=0.19\linewidth]{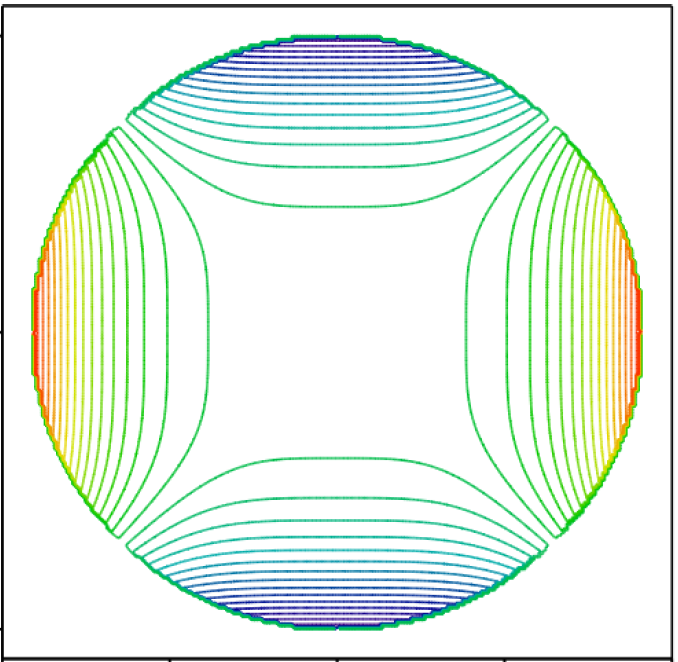}
    \includegraphics[width=0.19\linewidth]{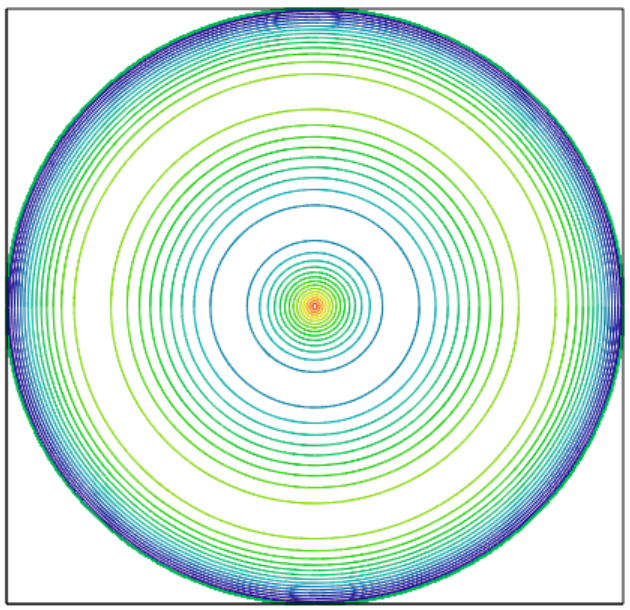}  
    \includegraphics[width=0.19\linewidth]{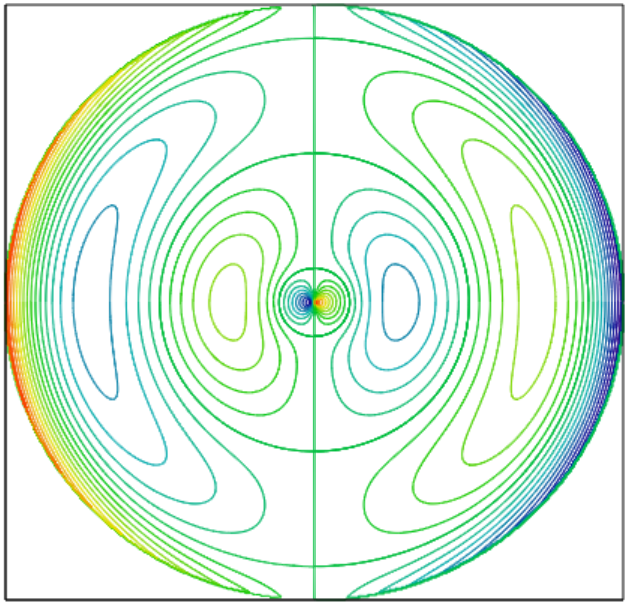}
    \includegraphics[width=0.19\linewidth]{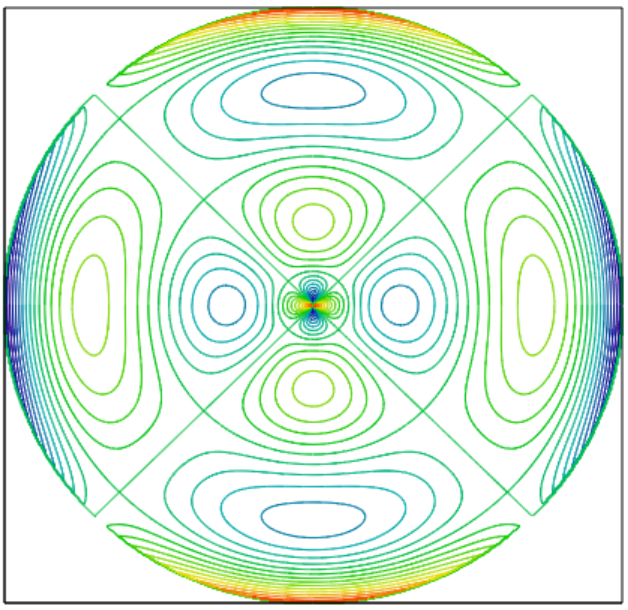}
    \includegraphics[width=0.19\linewidth]{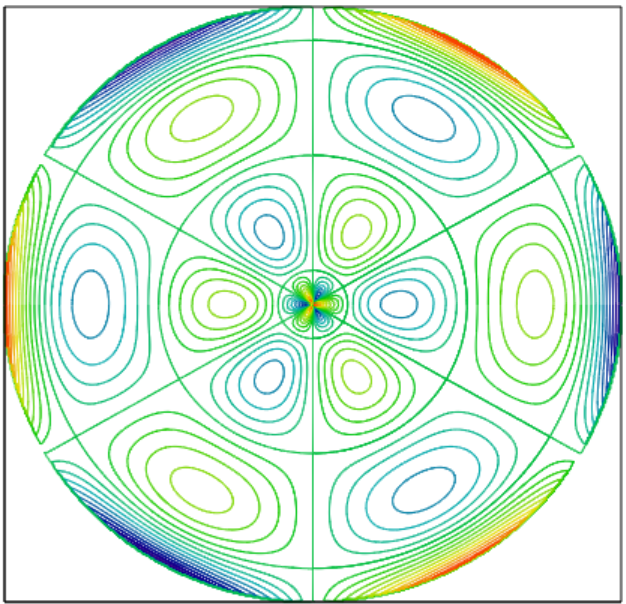}
\includegraphics[width=0.19\linewidth]{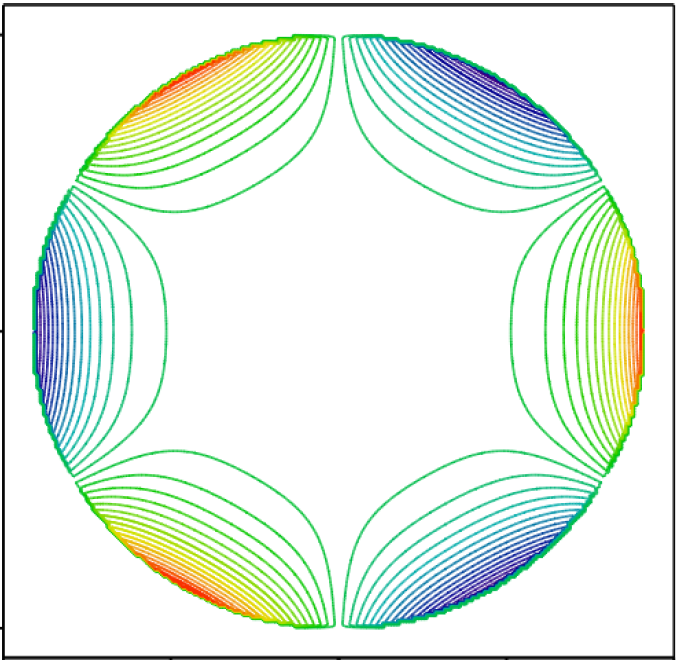}
    \caption{First four columns: the basis functions $fc_{n,m}(r,\phi)$ for $n$=0, 1, 2 and 3 (rows 1, 2, 3 and 4)
    and for $m$=0, 1, 2 and 3 (columns 1, 2, 3 and 4): 
    the color scale ranges from red ($\sqrt{\frac{2n+1}{\pi}}$) to green (0) to blue ($-\sqrt{\frac{2n+1}{\pi}}$), except for $fc_{n,0}$ which range from $\sqrt{\frac{2n+1}{2\pi}}$ to -$\sqrt{\frac{2n+1}{2\pi}}$.
    Fifth column: the basis functions for the PRM method for $m$=0, 1, 2 and 3 (rows 1, 2, 3 and 4).}
    \label{fig:basis}
\end{figure*}

When analysis of images is carried out in practise, only a finite number of radial and angular modes are considered. 
Images have finite resolution and have noise which means that modes which have finer scale than the image resolution or that are dominated by noise are not meaningful. 
For images with low noise and high resolution, this could mean retaining a large number of modes, such as $n$ and $m$ as large as $\sim$10 to 100. In practice, one might be interested only in low order structure: 
e.g. monopole ($n=0$ or $m=0$), dipole ($n=1$ or $m=1$), quadrupole ($n=2$ or $m=2$)  or octuupole ($n=3$ or $m=3$).

The cosine basis functions, Equation~\ref{eqn.3}, are illustrated in Figure~\ref{fig:basis} for $n,m$ =0, 1, 2 and 3. 
The sine basis functions are the same but rotated by 90$^\circ/m$. 
The incomplete PRM basis functions for $m$= 1, 2 and 3 were illustrated in Figure 1 of \cite{2019arXiv190911803R}, with $m=0$ uniform, the same as the right-most column (rows 2, 3, and 4) in Figure~\ref{fig:basis} here.
The PRM uses only functions with $m=n$ (i.e. incomplete) and with non-orthogonal radial functions instead of orthogonal radial functions.  
These are zero at the center of the image (except for $m$=0) and near-zero over an increasingly large area around the center as $m$ increases, so do not represent angular-dependent emission near the center of the image. 

\section{Supernova Remnant X-ray Image Sample}\label{sec:sample}

\cite{2011ApJ...732..114L} used the PRM method to analyze a set of 24 SNR images in the energy band 0.5-2.1 keV, dominated by thermal plasma emission. In addition, they analyzed a subset of 17 SNR images with significant emission in the 1.75-2.0 keV band, which has a strong contribution from \texttt{Si XIII} line emission.  
Here our goal is to test the new method, which uses a complete and orthogonal set of multipoles, on the main set of 24 images with the same energy band as 
the main set of 24 SNRs analyzed by \cite{2011ApJ...732..114L}.

\begin{figure*}
    \centering
    \includegraphics[width=\linewidth]{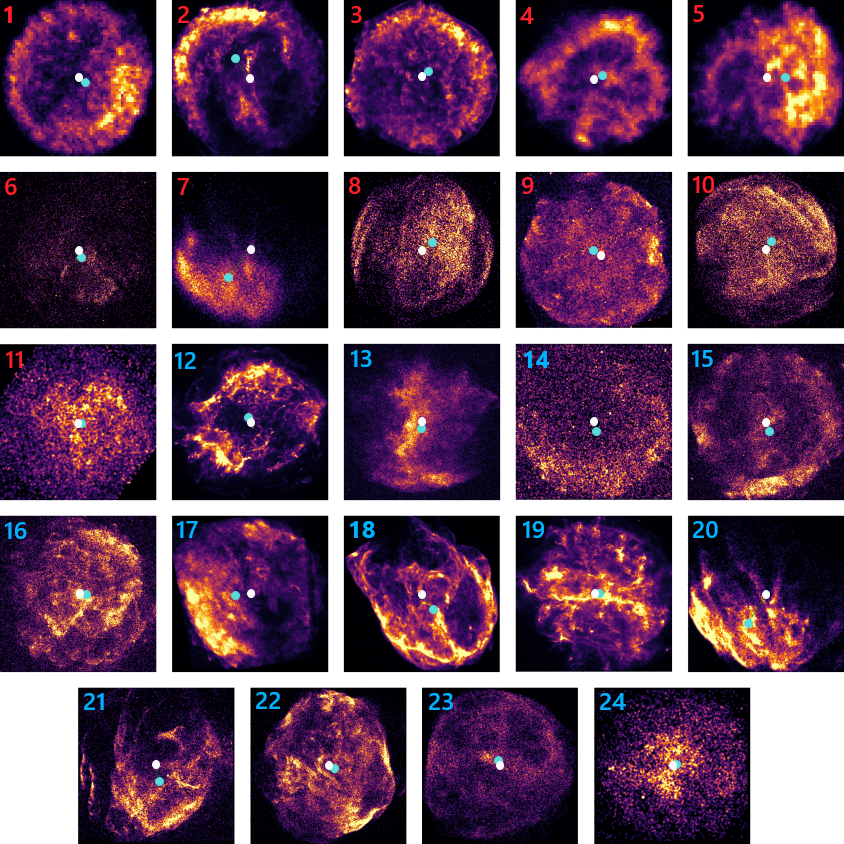}
    \caption{0.5–2.1 keV band X-ray images of the 24 SNRs used in this study. The numbers correspond to the SNRs in the Table \ref{tab1} and the red and blue numbers denote the Type Ia and CC SNRS, respectively. The physical center of the SNR is shown by a white dot and centroid by a cyan dot.}
    \label{fig:images}
\end{figure*}

\begin{table}[!ht]
    \centering
    \caption{Supernova Remnants with \textit{Chandra} X-ray Images}\label{tab1}
    \begin{tabular}{llll}
    \hline
        \# & SNR          & Observation ID(s)  \\ 
        \hline 
        Type Ia \\
    \hline
        1  & $0509-67.5$  & 776                \\    
        2  & Kepler       & 6714               \\ 
        3  & Tycho        & 3837, 10095, 15998 \\ 
        4  & $0519-69.0$  & 118                \\ 
        5  & N$103$B      & 125                \\ 
        6  & G$337.2-0.7$ & 2763               \\ 
        7  & DEM L$71$    & 775                \\ 
        8  & $0548-70.4$  & 1992               \\
        9  & G$272.2-3.2$ & 9147, 10572        \\ 
        10 & $0534-69.9$  & 1991               \\ 
        11 & G$344.7-0.1$ &  4651, 5336 \\ \\
        \hline
        Core-collapse\\
        \hline \\
        12 & CAS A        & 4634               \\         
        13 & W49B         & 117                \\ 
        14 & G$15.9+0.2$  & 5530               \\        
        15 & G$11.2-0.3$  & 780                \\ 
        16 & Kes 73       & 729                \\  
        17 & RCW 103      & 970                \\ 
        18 & N$132$D      & 5532               \\ 
        19 & G$292.0+1.8$ & 6677, 6679         \\         
        20 & $0506-68.0$  & 2762               \\
        21 & Kes 79       & 1982, 17453, 17659 \\         
        22 & N49B         & 1041               \\ 
        23 & B$0453-68.5$ & 1990               \\         
        24 & N206         & 3848               \\ \\
        \hline
    \end{tabular}
\end{table}

To create images for analysis in this project, the first \textit{Chandra} ACIS observation for each SNR from the list of the 24 SNRs in \cite{2011ApJ...732..114L} was obtained from the \textit{Chandra} observation archive. 
The first observation was chosen because the ACIS CCDs have significant age-degradation with time\footnote{see the \textit{Chandra} effective area viewer at  https://cxc.harvard.edu/cgi-bin/prop\_viewer/build \_viewer.cgi?ea.}. 
For SNRs which had ACIS chip gaps crossing the image, further observation IDs were included and combined to cover the chip gaps.
The 24 SNRs and their observation IDs are given in Table~\ref{tab1} with their SNR type.
The SNR G$344.7-0.1$ was listed as unclassified by \cite{2011ApJ...732..114L}, but has since been classified as Type Ia \citep{2020ApJ...897...62F}.

The images were created using the \texttt{\textit{Chandra} Interactive Analysis of Observations (CIAO)} software. 
This includes \texttt{chandra\_repro}, which tells \texttt{CIAO} to reprocess the raw observation data with the latest calibrations.  
\texttt{fluximage} was used to create exposure-corrected flux images from the reprocessed observations. 
The lower and upper limits of photon energy used to create the flux images were 0.5 and 2.1 keV to match the energy range used by \cite{2011ApJ...732..114L} for their 24-SNR sample.  

\section{Multipoles for the SNR Images}\label{sec:SNRmult}

The 2-D multipole analysis is defined using a circular boundary. 
Thus there is the question of how to choose the center and radius of the circle. 
In practise, one first subtracts the background level from the image and removes any unrelated point sources, leaving only the SNR emission in the image.
One option to obtain the circular boundary is to use 
the minimum-sized circle which encompasses all emission from the SNR. 
This defines the center, which we call ``physical center" and the radius of the circle.
Another option is to find the brightness centroid of the SNR then find the circle centered on this centroid which is large enough to encompass all emission from the SNR.
Using the centroid as center yields a radius larger than for the first case and it necessarily includes area with no SNR emission inside this larger circle.
Figure~\ref{fig:images} shows the physical center (by a white dot) and the centroid (by a cyan dot) for each SNR.
Here we give results for both choices of circle: physical center and centroid. 
However we consider the first option as a more physical choice.

For the \textit{Chandra} X-ray images the background levels were  small compared to the SNR brightness. 
The mean background brightness and SNR mean brightness were measured for all 24 SNR images.
This yielded ratios of background to SNR mean brightness of 0.001 to 0.1 with a mean ratio of 0.04 and standard deviation of values of 0.04.
This simplified the selection of the outer circular boundary of the SNR emission: we chose the minimum bounding circle of the SNR emission. 
The background emissions of the \textit{Chandra} images inside the circle were faint compared to the SNR emission. 

For both cases (physical center and centroid), the circle (or aperture) radius ($R_{\textrm{ap}}$) was found in pixels.
The polar coordinates of the pixels were calculated for the two cases, then the radius values were scaled so that $r=1$ at the circle boundary for the two cases. 
The analysis is limited to the lowest order modes (n= 0, 1, 2, 3 and m= 0, 1, 2, 3). 
I.e. the sums in Equations~\ref{eqn.9} and \ref{eqn.10} have an upper limit of 3.
We use Equations~\ref{eqn.2}–\ref{eqn.10} to calculate the radial and angular multipoles of the X-ray SNRs listed in Table \ref{tab1}. 

We calculated the statistical noise in the multipole powers for each SNR as follows.
We created multiple realizations of each SNR image, each generated with noise added to each pixel in the image using a Gaussian random generator with Gaussian mean and standard deviation based on the background level for each SNR image. 
Then we calculated the multipole powers for each realization of a given SNR and used the standard deviation of these multipole powers as the noise in the multipole powers. 
Rather than listing the noise of multipole powers for each of the 24 images, which were similar, in Table~\ref{tab:center} we give the means and standard deviations for the set of Type Ia SNRs and for the set of Type CC SNRs.
The errors in the multipole powers from noise are small compared to the difference in multipole powers between SNRs: the largest noise is 0.004 compared to the difference in a given multipole power between SNRs of 0.01 to 0.5.
The noise values for the case of using the centroid are essentially the same, so the noise values are not repeated in Table~\ref{tab:centroid}.

An estimate of systematic errors in the multipole powers was also calculated using two tests: offsetting the center of the SNR from its best center and offsetting the radius from its best radius.
The center and radius are determined by manually placing the smallest circle that bounds the SNR emission, then finding the center and radius of that circle. 
Tests indicate that this process is accurate to the larger of 1 pixel or $\simeq$1\% of the SNR radius.
For the first test the centers of the SNRs were offset by the larger of 1 pixel or 1\% of the SNR radius, and the SNR multipole powers were recalculated. 
For the second test the radius of the SNRs were increased or decreased by the larger of 1 pixel or 1\% of the SNR radius, and the SNR multipole powers were recalculated.
For offset in the centers, the SNRs with radius larger than $\sim100$ pixels 
had their multipole powers change by 0.0001 to 0.02 (mean 0.004).
The SNRs with radius smaller than $\sim100$ pixels had their multipole powers change by 0.0005 to 0.04 (mean 0.02).
For change in the radius, the SNRs with radius larger than $\sim100$ pixels 
had their multipole powers change by 0.00003 to 0.007 (mean 0.003).
The SNRs with radius smaller than $\sim100$ pixels had their multipole powers change by 0.001 to 0.04 (mean 0.01).
In summary, the systematic errors in the calculated multipole powers are typically less than $\simeq$0.02, as detailed above.

\begin{table*}[h!]
\begin{center}
\caption{Radial and Angular Power-Ratios using Physical Center$^a$}
   \centering
    \begin{tabular}{c l c c c c c c c c c}
    \hline
\# & SNR & $P_{rad,0}$ & $P_{rad,1}$ & $P_{rad,2}$ & $P_{rad,3}$ & $P_{ang,0}$ & $P_{ang,1}$ & $P_{ang,2}$ & $P_{ang,3}$    \\ 
\hline
\multicolumn{11}{c}{Type Ia} \\ 
\hline
1 & $0509-67.5$ 
&1.127 &0.088 &0.110 &0.042 &1.157& 0.150 &0.057& 0.003 \\ 
2 & Kepler & 
1.167 &0.163 &0.071 &0.036 &1.040 &0.306 &0.078 &0.013\\
3 & Tycho & 
 1.035 &0.040 &0.038& 0.091 &1.108& 0.077& 0.012&0.007\\
4 & $0519-69.0$ & 
1.039 &0.014 &0.063 &0.047 &1.073 &0.047 &0.010& 0.034\\ 
5 & N103B & 
1.232& 0.027& 0.059 &0.023 &1.052 &0.239 &0.037& 0.014\\ 
6 & G$337.2-0.7$ & 
1.110 &0.137& 0.037 &0.014& 1.134& 0.099& 0.036 &0.029\\
7 & DEM L71 & 
1.694& 0.110 &0.078 &0.026 &1.026 &0.714 &0.146 &0.023\\ 
8 & $0548-70.4$ & 
1.139 &0.127 &0.057& 0.015 &1.146& 0.144& 0.033& 0.016\\
9 & G$272.2-3.2$ & 
1.010& 0.026& 0.002& 0.008& 1.017& 0.017& 0.008& 0.003\\
10 & $0534-69.9$ & 
1.020 &0.079 &0.023 &0.030& 1.073 &0.056& 0.017 &0.006\\
11 & G$344.7-0.2$ & 
1.010& 0.135& 0.015& 0.009& 1.133& 0.017& 0.008& 0.012\\
\vspace{2pt}\\
\multicolumn{2}{l}{MeanNoise$_{Type Ia}^b$} &	0.0008&	0.002&	0.0004&	0.0004&	0.002&	0.001&	0.0003& 0.0002\\
\multicolumn{2}{l}{StDevNoise$_{Type Ia}^b$}  & 	0.0007&	0.004&	0.0004&	0.0005&	0.004&	0.0008&	0.0003& 0.0003\\
\vspace{2pt}\\
\multicolumn{2}{l}{Mean$_{Type Ia}$} & 1.144&	0.086&	0.050&	0.031&	1.087&	0.169&	0.040&	0.015& \\
\multicolumn{2}{l}{StDev$_{Type Ia}$}  & 0.197&	0.053&	0.031&	0.024&	0.051&	0.202&	0.042&	0.010& \\
\hline
\multicolumn{11}{c}{Type CC} \\ 
\hline
12 & CAS A & 
1.109& 0.017 &0.294& 0.046& 1.231& 0.071& 0.1193& 0.045\\
13& W49B & 
1.063 &0.246& 0.0199& 0.015& 1.217& 0.050& 0.050& 0.027\\
14 & G$15.9+0.2$ & 
1.049 &0.068& 0.009 &0.011& 1.054& 0.069& 0.010& 0.004\\ 
15 & G$11.2-0.3$ & 
1.056 &0.032& 0.075 &0.075& 1.142 &0.057 &0.019& 0.019\\
16 & Kes 73 & 
1.036 &0.245 &0.086 &0.132 &1.424& 0.057 &0.012& 0.004\\
17 & RCW 103 & 
1.229& 0.022& 0.249& 0.021& 1.158& 0.266& 0.092& 0.005 \\
18 & N132D & 
1.105& 0.067& 0.077 &0.035 &1.024& 0.156& 0.067& 0.037\\
19 & G$292.0+1.8$ &
1.072&0.195 &0.023 &0.005& 1.170& 0.034& 0.072& 0.022\\ 
20 & $0506-68.0$ & 
1.580& 0.155& 0.059 &0.011 &1.009 &0.675 &0.096& 0.025\\
21 & Kes79 & 
1.212& 0.020 &0.125& 0.035& 1.092& 0.258 &0.021& 0.020\\
22 & N49B & 
1.022& 0.070& 0.039& 0.017& 1.064& 0.031& 0.032 &0.021\\
23 & B$0453-685$ & 
1.037& 0.060& 0.009& 0.009& 1.053& 0.030& 0.025 &0.007\\
24 & N206 & 
1.009 &0.339& 0.015& 0.033 &1.339 &0.006 &0.037& 0.014\\
\vspace{2pt}\\
\multicolumn{2}{l}{MeanNoise$_{Type CC}^b$} &	0.002&	0.001&	0.0004&	0.0001&	0.0004&	0.003&	0.0007& 0.0002\\
\multicolumn{2}{l}{StDevNoise$_{Type CC}^b$}  & 	0.008&	0.002&	0.001&	0.0002&	0.001&	0.009&	0.002& 0.0004\\
\vspace{2pt}\\
\multicolumn{2}{l}{Mean$_{Type CC}$} & 1.121&	0.118&	0.083&	0.034&	1.152&	0.135&	0.050&	0.019& \\
\multicolumn{2}{l}{StDev$_{Type CC}$}  &0.154&	0.106&	0.091&	0.035&	0.125&	0.183&	0.036&	0.012 & \\
\hline
\vspace{2pt}\\
\multicolumn{2}{l}{Mean$_{Type Ia}$-Mean$_{Type CC}$} & 0.022&	-0.032&	-0.033&	-0.003&	-0.065&	0.034&	-0.010&	-0.004& \\
\multicolumn{2}{l}{StDev$_{Combined}$}  & 0.171&	0.086&	0.071&	0.030&	0.102&	0.188&	0.038&	0.011& \\
\hline
    \end{tabular}
    \label{tab:center}
\end{center}
\tablecomments{$^{\textrm{a}}$: StDev stands for standard deviation.}
\tablecomments{$^{\textrm{b}}$: Mean Noise and StDev Noise are the mean and standard deviation in the resulting power ratios caused by image statistical noise.}
\end{table*}

\begin{table*}[h!]
\begin{center}
\caption{Radial and Angular Power-Ratios using Centroid$^a$}
   \centering
    \begin{tabular}{c l c c c c c c c c c c}
    \hline
\# & SNR & $P_{rad,0}$ & $P_{rad,1}$ & $P_{rad,2}$ & $P_{rad,3}$ & $P_{ang,0}$ & $P_{ang,1}$ & $P_{ang,2}$ & $P_{ang,3}$ & Centroid Offset \\ 
& & & & & & & & &  & (units of R)  \\ 
\hline
\multicolumn{12}{c}{Type Ia} \\ 
\hline
1 & $0509-67$.5 & 
1.088 &0.093 &0.218& 0.064& 1.172& 0.240& 0.037& 0.015& 0.112\\
2 & Kepler & 
1.039 &0.349 &0.213 &0.299 &1.646 &0.193 &0.032 &0.030  & 0.314 \\
3 & Tycho & 
1.031 &0.067 &0.229 &0.038& 1.235& 0.117& 0.009& 0.004& 0.091\\
4 & $0519-69.0$ & 
1.008 &0.164& 0.092 &0.075& 1.236& 0.018& 0.031& 0.055& 0.105 \\
5 & N103B & 
1.037 &0.443& 0.079& 0.107& 1.490& 0.110 &0.031 &0.034 & 0.255 \\
6 & G$337.2-0.7$ & 
1.047& 0.264 &0.044 &0.002& 1.237 &0.051 &0.052& 0.016  & 0.089 \\
7 & DEM L71 & 
1.055 &1.036 &0.112 &0.056& 2.098& 0.082& 0.062& 0.017& 0.445 \\
8 & $0548-70.4$ & 
1.037& 0.436 &0.023 &0.021 &1.406& 0.084& 0.006 &0.021& 0.143 \\
9 & G$272.2-3.2$ & 
1.002&	0.070&	0.022&	0.027&	1.101&	0.012&	0.004&	0.004 & 0.098\\
10 & $0534-69.9$ & 
1.011 &0.222 &0.066& 0.008& 1.276& 0.008 &0.008& 0.014& 0.123 \\
11 & G$344.7-0.2$ & 
1.006&	0.194&	0.015&	0.006&	1.193&	0.015&	0.005&	0.008 & 0.047 \\
\vspace{2pt}\\
\multicolumn{2}{l}{Mean$_{Type Ia}$}  & 1.033&	0.303&	0.101&	0.064&	1.372&	0.085&	0.025&	0.020&0.166\\
\multicolumn{2}{l}{StDev$_{Type Ia}$}  & 0.026&	0.277&	0.082&	0.085&	0.288&	0.077&	0.020&	0.015 &0.121\\
\hline
\multicolumn{12}{c}{Type CC} \\ 
\hline
12 & CAS A & 
1.082& 0.328& 0.222& 0.208& 1.584& 0.077& 0.152& 0.026& 0.069 \\
13 & W49B & 
1.050 &0.396 &0.035& 0.022 &1.355& 0.079&0.051 &0.017& 0.074 \\
14 & G$15.9+0.2$ & 
1.018 &0.010 &0.160 &0.039 &1.135& 0.066 &0.020& 0.006 & 0.120\\
15 & G$11.2-0.3$ & 
1.025 &0.098& 0.096& 0.033 &1.148& 0.060& 0.033&0.012& 0.130 \\
16 & Kes 73 & 
1.027 &0.305& 0.060& 0.037& 1.283& 0.108& 0.015& 0.024& 0.094 \\
17 & RCW 103 & 
1.116&	0.381&	0.063&	0.055&	1.315&	0.248&	0.035&	0.018 & 0.215 \\
18 & N132D & 
1.030 &0.299& 0.084& 0.054& 1.323& 0.078& 0.046& 0.020& 0.247 \\
19 & G$292.0+1.8$ & 
 1.054& 0.383 &0.029 &0.019& 1.353& 0.054 &0.057 &0.022& 0.057\\
20 & 0$506-68.0$ & 
1.048 &0.882 &0.091 &0.059 &1.902& 0.085& 0.066& 0.027& 0.427 \\
21 & Kes79 & 
1.078 &0.339& 0.098& 0.109& 1.436 &0.137 &0.043 &0.008& 0.211 \\
22 & N49B & 
1.020 &0.113 &0.052 &0.022 &1.136& 0.018& 0.025& 0.029& 0.069 \\
23 & B$0453-685$ & 
1.017 &0.203 &0.015& 0.030 &1.219 &0.016& 0.024& 0.007& 0.072 \\
24 & N206 & 
 1.015& 0.428& 0.048& 0.081 &1.434& 0.025&0.073& 0.040& 0.044 \\
\vspace{2pt}\\
\multicolumn{2}{l}{Mean$_{Type CC}$} & 1.045&	0.320&	0.081&	0.059&	1.356&	0.081&	0.049& 0.020&0.141 &\\
\multicolumn{2}{l}{StDev$_{Type CC}$} & 0.031&	0.213&	0.057&	0.051&	0.210&	0.061&	0.036&	0.010& 0.109
& \\
\hline
\vspace{2pt}\\
\multicolumn{2}{l}{Mean$_{Type Ia}$-Mean$_{Type CC}$} &  -0.012&	-0.017&	0.020&	0.005&	0.016&	0.004&	-0.024&	0.000&0.025& \\
\multicolumn{2}{l}{StDev$_{Combined}$}  & 0.029&	0.239&	0.069&	0.067&	0.243&	0.067&	0.032&	0.012&0.113& \\
\hline
    \end{tabular}
    \label{tab:centroid}
\end{center}
\tablecomments{$^{\textrm{a}}$: StDev stands for standard deviation.}
\end{table*}

The cosine and sine multipole coefficients $ac_{n,m}$ ($n=0$ to 3 and $m=0$ to 3) and $as_{n,m}$ ($n=0$ to 3 and for $m=1$ to 3) for the 21 SNR images were calculated; a total of 28 coefficients for each image.
We give the radial and angular power ratios here (Equations~\ref{eqn.9} and \ref{eqn.10}, a total of 8 values for each image) to summarize the results. 

Table~2 lists the power ratios and size of each SNR in image pixels ($R_{ap}$) for the case of using the physical center as image center. 
It also gives the means and standard deviations for the power ratios for each SNR type (Type Ia and Type CC).
For each of the 8 power-ratios (4 radial and 4 angular) the mean for Type Ia and the mean for Type CC are within 1 standard deviation of each other, and often differ by much less than a standard deviation.  

Table~3 lists the power ratios and sizes for the case of using the centroid as image center, and also lists the offset between the physical center and the centroid.
Again, for each of the 8 power-ratios the mean for Type Ia and the mean for Type CC are within 1 standard deviation of each other.
The centroid offsets vary significantly between different SNRs (from 0.044 for N206, CC Type, to 0.445 for DEM L71, Type Ia) but the mean centroid offsets for Type Ia and Type CC are similar, well within the standard deviations.

We also computed all 16 power ratios $P(n,m)$, $n, m=0,1,2,3$ for each SNR.
The means and standard deviations were computed for Type Ia SNRs and Type CC SNRs.
For each of the 16 $P(n,m)$, the mean for Type Ia and mean for Type CC were equal within one standard deviation. 

The choice of physical center is compared to choice of centroid for the multipole power ratios results.
As noted above, when using the centroid as center one requires a larger radius circle ($R_{ap}$) to encompass the emission from the SNR.
This results in adding a region with no SNR emission for the centroid case which is inside the circle but outside the SNR emission. 

The effect on the power ratios is as follows.
$P_{\text{rad,0}}$ is smaller for the centroid case, which is expected because of the larger area of no SNR emission inside the analysis circle.
$P_{\text{rad,1}}$ is larger for the centroid case, because it includes the large $P(1,0)$ which is caused by offset of the center (e.g. see the image for Kepler (2) in Figure~\ref{fig:images}) which increases linear ($n=1$) radial dependence.
$P_{\text{ang,0}}$ is larger for the centroid case, because it is the sum of $P(n,0)$ over $n$, thus includes the large $P(1,0)$ term.
The other power ratios not so different between center and centroid case. 
Thus the choice of center clearly affects the power ratios.
 
$P_{\text{ang,1}}$ is not 0 for the centroid case for two reasons. 
First, $P_{\text{ang,1}}$ is the sum of power ratios over $n$ for $m=1$,
Second, $P(1,1)$ is non-zero for the centroid case, as follows.
The centroid ($x_c ,y_c$) is found using $x_c=\sum_{pixels} x_i F_i$ and $y_c=\sum_{pixels} y_i F_i$ with $F_i$ the brightness of pixel $i$. 
In polar coordinates this is: $r_c cos(\phi_c)=\sum_{pixels} r_i cos(\phi_i) F_i$ and $r_c sin(\phi_c)=\sum_{pixels} r_i sin(\phi_i) F_i$.
$ac_{1,1}$ and $ac_{1,1}$ involve the integral (Equations (6) and (7), or equivalently the Riemann sum) of $P_1^*(r_i) cos(\phi_i) F_i$ and $P_1^*(r_i) sin(\phi_i) F_i$. 
With $P_1^*(r)=2r-1$, this integral does not vanish (which it would if $P_1^*(r)$ was replaced by $r$).

\begin{figure*}
    \centering
    \includegraphics[width=\linewidth]{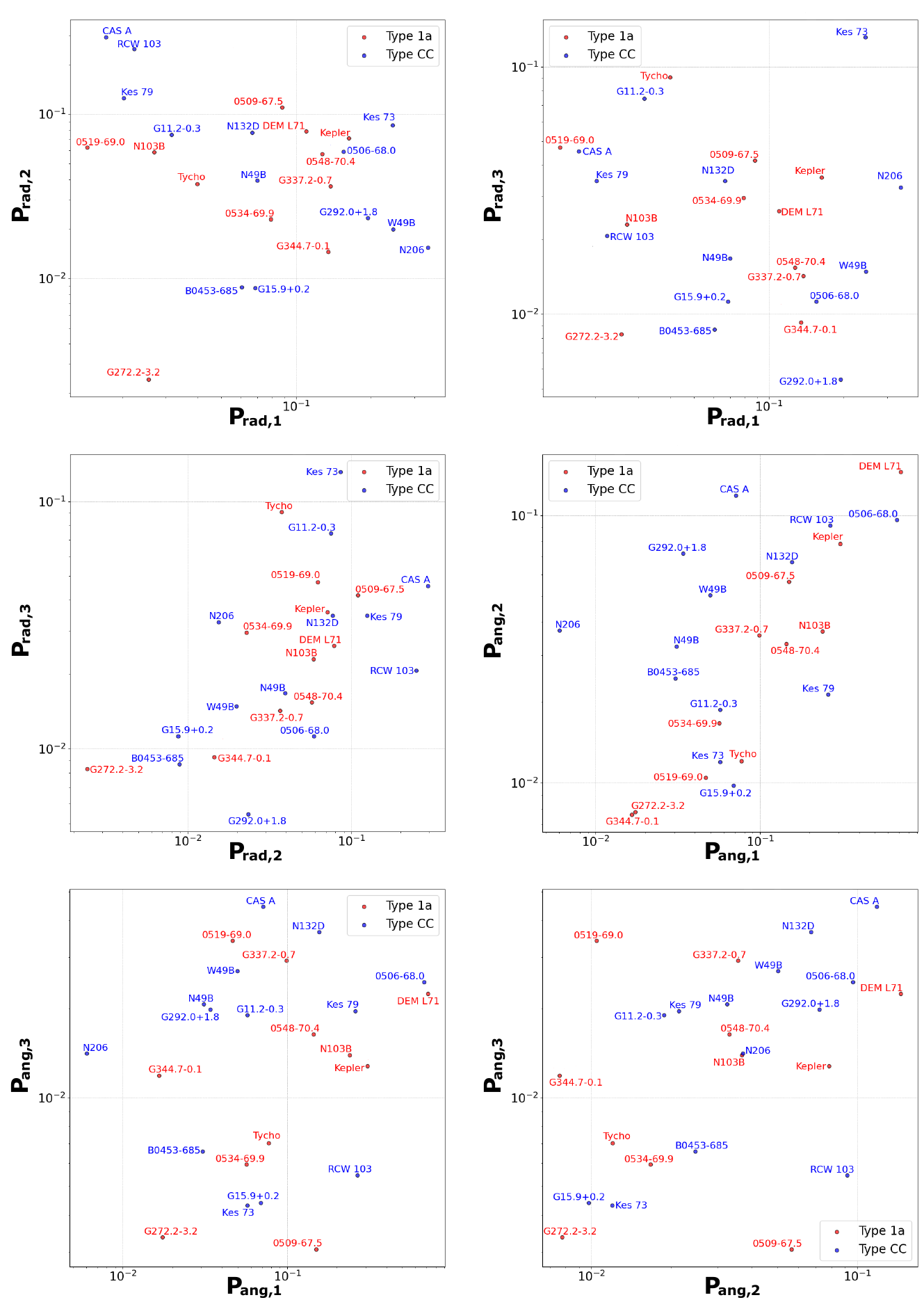}
    \caption{Power ratio plots for SNRs of known Type Ia (red) and CC SNe (blue) where the physical center of the SNR is taken to be the analysis  center.}
    \label{fig:newMultphyscen}
\end{figure*}

\begin{figure*}
    \centering
    \includegraphics[width=\linewidth]{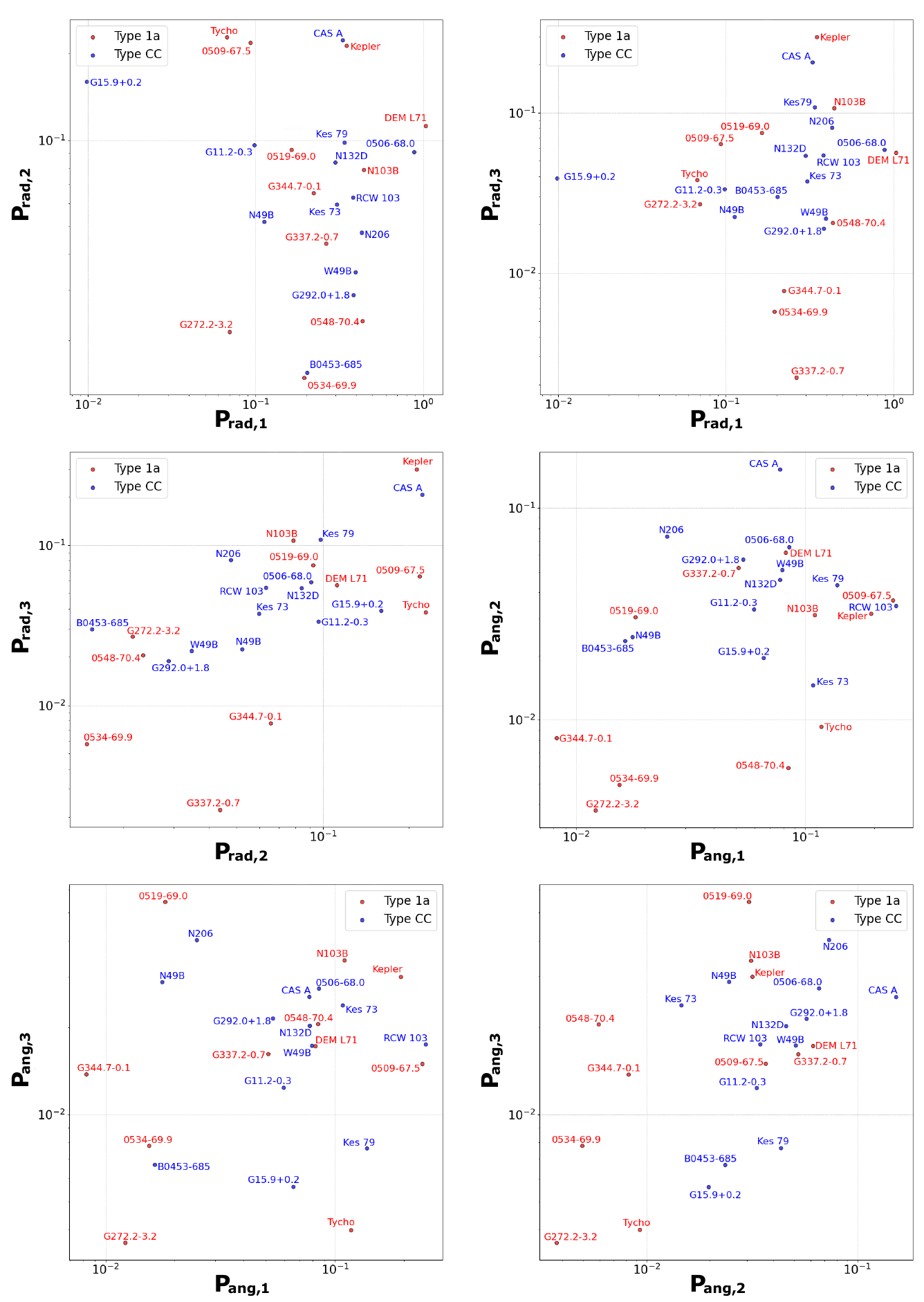}
    \caption{Power ratio plots for SNRs of known Type Ia (red) and CC SNe (blue) where the centroid of the SNR is taken to be the analysis center.}
    \label{fig:newMultcentroid}
\end{figure*}

Figure~\ref{fig:newMultphyscen} shows the plots of one power ratio vs. another for the physical center case (excluding $P_{rad,0}$ and $P_{ang,0}$),
resulting in 3 plots for the different combinations of $P_{rad,1}$, $P_{rad,2}$ and $P_{rad,3}$ and 3 plots for the different combinations of $P_{ang,1}$, $P_{ang,2}$ and $P_{ang,3}$.
The two types of SNRs (Type Ia in red, Type CC in blue) are scattered and mixed in all six diagrams.
One difference between Type Ia and Type CC is that Type CC have larger scatter in $P_{rad,2}$ than Type Ia.
This can be seen in Table~\ref{tab:center} where the standard deviation of $P_{rad,2}$ is 0.091 for Type CC vs. 0.031 for Type Ia. 
For the 3 plots involving $P_{ang,m}$, the Type Ia and Type CC appear more uniformly mixed. 
The Type CC sample has more SNRs with low $P_{ang,1}$ than Type Ia, although the means are still different by less than the standard deviations.

Figure~\ref{fig:newMultcentroid} shows the plots of one power ratio vs. another for the case of centroid as analysis centre.
For the radial power ratios ($n=$1, 2, 3; the first 3 panels), the Type Ia and Type CC are well mixed.
For the angular power ratios $m=$1, 2, 3; the last 3 panels), the Type Ia and Type CC are well mixed, except that there are more Type Ia SNRs with small $P_{ang,2}$ than Type CC SNRs.
Overall, Type Ia and Type CC are fairly well mixed in these diagrams, similar to the case where the physical center is used as analysis center.

\begin{figure}
    \centering
    \includegraphics[width=\linewidth]{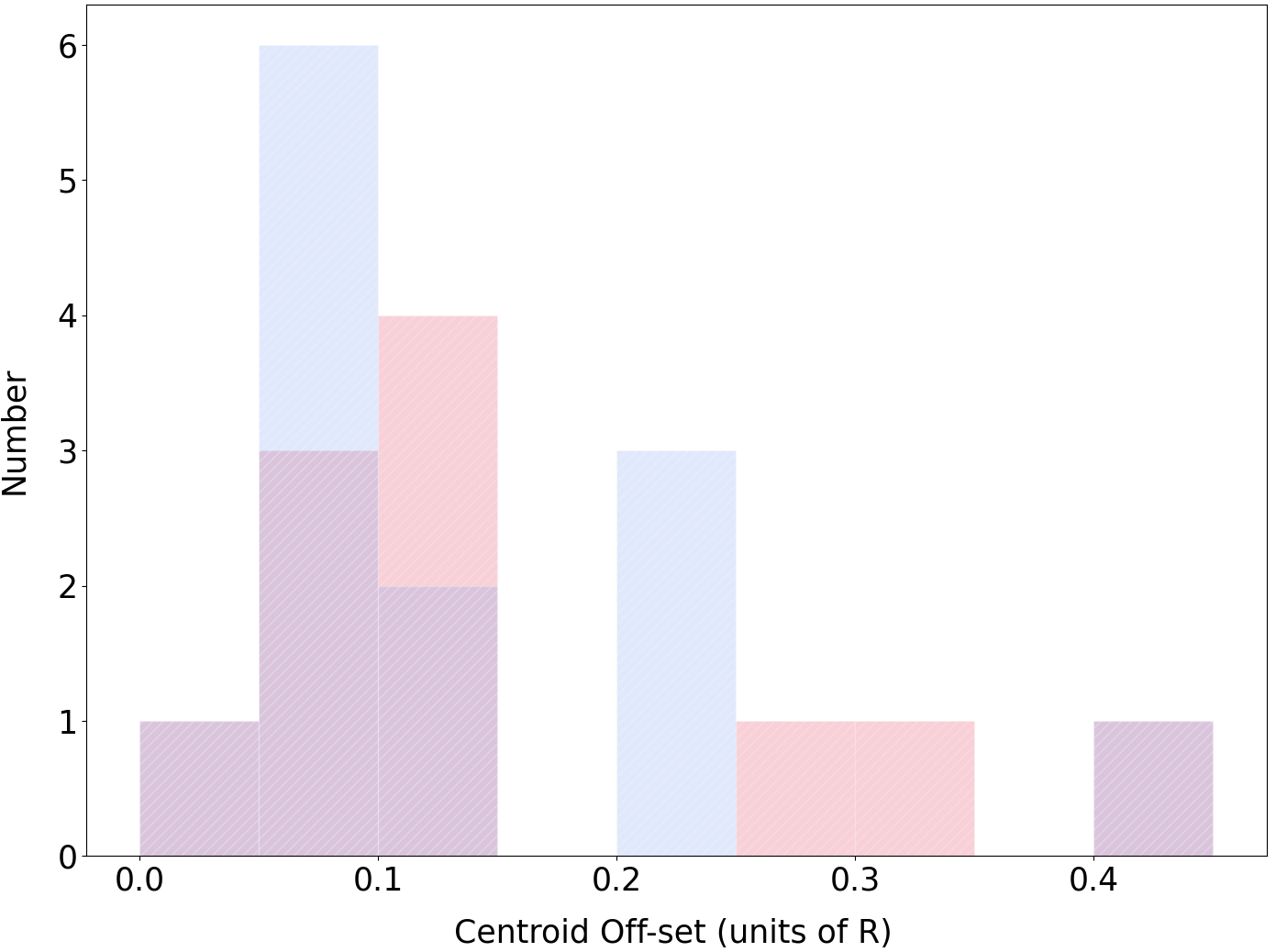}
    \caption{Histogram of centroid offsets, in units of SNR radius, for remnants of known Type Ia (red) and CC  (blue). The mean offsets for Type Ia and CC are not significantly different.}
    \label{fig:centroid}
\end{figure}

We show the distributions of centroid offsets in Figure~\ref{fig:centroid}.
This is a separate test of asymmetry that can be used compare Type Ia and Type CC SNRs. 
The difference in mean centroid offset is much smaller than the standard deviation of either Type Ia or Type CC SNRs, thus the two types are not distinguishable in their centroid offsets.

\section{Discussion}\label{sec:disc}

\subsection{Results of the current work}\label{sec:discNew}

From Section~\ref{sec:SNRmult}, the results of the multipole analysis of the SNR images yields the following. 
We consider using the SNR center as the center for analysis to be the better choice than using the image centroid.
The Type Ia and CC type SNRs have similar means of the 8 different power-ratios considered: $P_{rad,n}$, $n=0$ to 3 and $P_{ang,m}$, $m=0$ to 3.
In the diagrams (Figure~\ref{fig:newMultphyscen}) of one power-ratio vs. another for radial power-ratios (3 diagrams) and for angular power-ratios (3 diagrams), the Type Ia and Type CC SNRs are mixed.
There is a wider distribution of $P_{rad,2}$ for Type CC than for Type Ia,
and $P_{ang,1}$ has more Type CC at low values than Type Ia. 
In addition to the radial and angular power ratio diagrams shown in Figures~\ref{fig:newMultphyscen} and \ref{fig:newMultcentroid}, we considered all combinations of $P(n,m)$ and $P(n',m')$ for all $n,m,n',m'$ combinations, and did not find a clear distinction in any of them between Type Ia and Type CC SNRs. 

\subsection{Comparison with Previous Work}

A multipole analysis was applied to X-ray images of SNRs by \cite{2011ApJ...732..114L}, using an incomplete and non-orthogonal set of basis functions. 
They analysed 24 images, including 10 Type Ia, 13 CC SNRs and one other. 
That work found a distinction between the Type Ia and CC type SNRs in the quadrupole vs. octupole ratio diagram, but did not find a difference between the two types when applying a wavelet analysis. 
The PRM analysis (using the same set of non-orthogonal incomplete basis functions) was applied to radio images of SNRs by \cite{2019arXiv190911803R}.
That work utilized 1420 MHz radio images of 5 Type Ia and 14 core-collapse SNRs. 
No clear distinction between the two types of SNRs was found in the quadrupole vs. octupole ratio diagram. 

The basis functions used in the above two studies had the issues that they are not a complete set of basis functions, nor are they orthogonal basis functions, as described in the  Sec.~\ref{complete}. 

The normalization used by \cite{2011ApJ...732..114L} to define the quadrupole and octupole ratios is unusual:
the $P_2$ and $P_3$ coefficients have dimensions of flux$^2$  (surface brightness times solid angle)$^2$, 
whereas the $P_0$ coefficient has dimensions of flux$^2$ times $(ln(R))^2$. 
This means that a factor in the location of each SNR in the $P_2/P_0$ vs. $P_3/P_0$ diagram is the radius of that SNR\footnote{Another unusual aspect is that the value of $ln(R)$ depends on the units chosen for $R$.}. 
A more logical approach is to normalize the whole SNR sample to the same radius. 
This was done by \cite{2019arXiv190911803R}, by rescaling the radius of each SNR to $e$ yielding $ln(R)=1$ for all SNRs in the sample.

\begin{figure*}
    \centering
    \includegraphics[width=\linewidth]{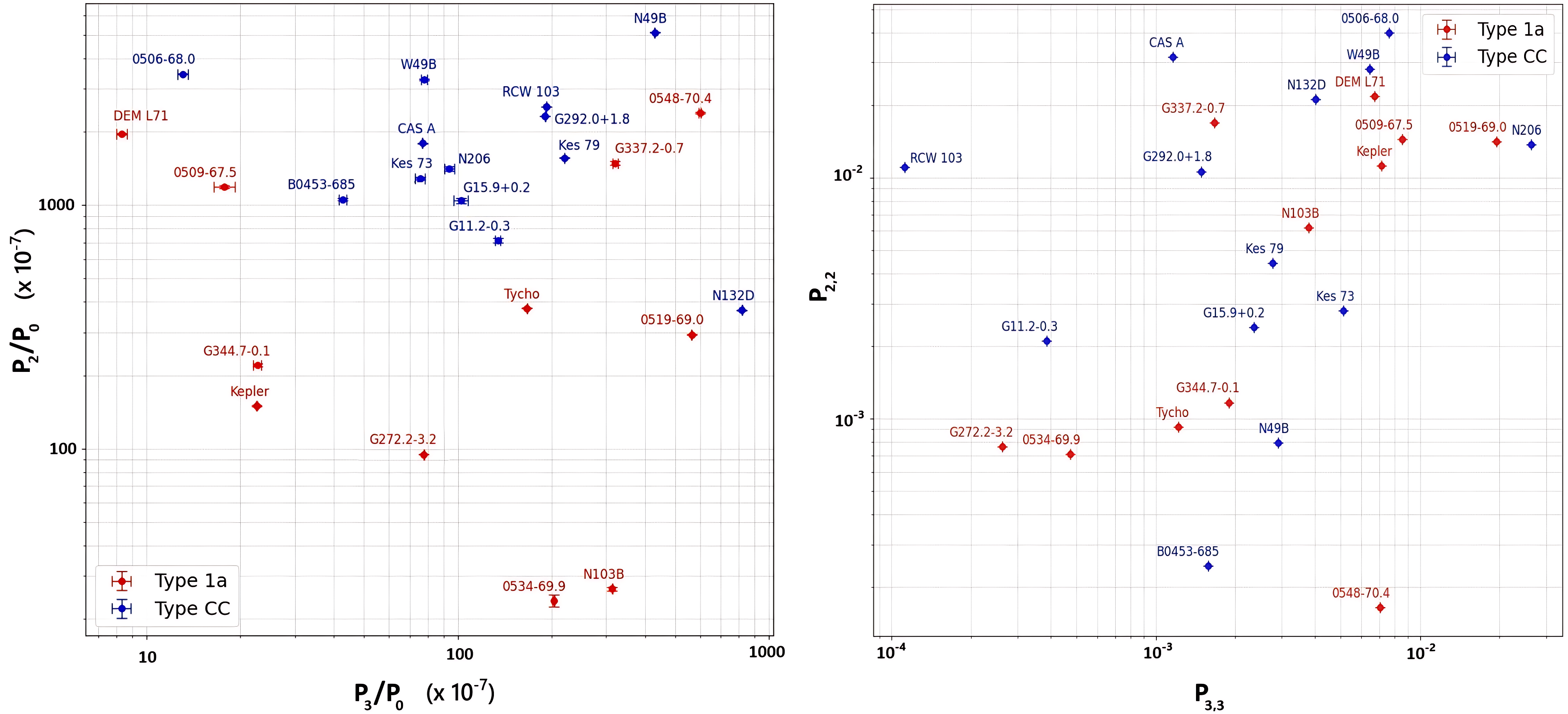}
    \caption{Left Panel: Power ratio diagram using the PRM method and centroid as center of analysis, which uses an incomplete and non-orthogonal set of basis functions, applied to X-ray images of Type Ia and Type CC SNRs.
    Right Panel: the equivalent diagram using orthonormal basis functions with centroid as center of analysis.}
    \label{fig:oldXRmult}
\end{figure*}

Thus we carried out a re-analysis of X-ray images, using the incomplete and non-orthogonal basis functions of the PRM, with the radius of each SNR scaled to $e$. 
The X-ray images consisted of the 24 listed in Table~\ref{tab1}, same as used by \cite{2011ApJ...732..114L}.
The results of calculating the $P_2/P_0$ vs. $P_3/P_0$ diagram are shown in the left panel of Figure~\ref{fig:oldXRmult}.
The Type Ia generally have smaller $P_2/P_0$ than CC SNRs, and the two types have essentially the same distribution of $P_3/P_0$, although
there is not a clear dividing line between the two types.
If we use a horizontal line at $P_2/P_0=500\times10^{-7}$ then 4 of the 11 Type Ia SNRs have $P_2/P_0$ above the line, like most Type CC values, and 1 of the 13 CC SNRs has $P_2/P_0$ below the line, similar to Type Ia values.
We note that \cite{2011ApJ...732..114L} concluded that there was a difference between Type Ia and CC images using the PRM method, but no distinction between based on their wavelet analysis.

As a further test, we show the equivalent of the $P_2/P_0$ vs. $P_3/P_0$ diagram using centroids for image centers but using orthonormal functions (the $P(2,2)$ vs. $P(3,3)$ diagram) in the right hand panel of Figure~\ref{fig:oldXRmult}. 
In this diagram, there is no clear distinction between Type Ia and Type CC SNRs.
This indicates that the unusual choice of functions of the PRM method is useful for (non-uniquely) separating Type Ia from Type CC SNRs, despite the fact that they are not good basis functions.
 
\section{Conclusions}

In the current study, we present an analysis method for multipole analysis of 2D images.
This method uses a complete and orthogonal set of basis functions and distinguishes between radial order $n$ and angular order $m$: a 2D set of basis functions is required for fully representing 2D images. 
The method is applied to a sample of 24 X-ray images of SNRs in the energy band 0.5-2.1 keV, which is dominated by thermal plasma emission.
Two cases are considered: one where the analysis center is the physical center of the SNR and one where the center is the centroid of the brightness distribution.
The brightness centroid offsets from the physical centers show no significant difference between Type Ia SNRs and Type CC SNRs.

From the multipole analysis, no clear distinction if found between X-ray images of Type Ia and of CC type SNRs in low order (monopole through octopole) radial or angular moments. 
We test the PRM method with a correction to remove the radius dependence and find that Type Ia SNRs have low $P_2/P_0$ values and Type CC have high $P_2/P_0$ in agreement with \cite{2011ApJ...732..114L}.
However, for the PRM method, 4 of the 11 Type Ia have high $P_2/P_0$ values and 1 of the 13 Type CC has a low $P_2/P_0$ value, meaning $P_2/P_0$ does not uniquely separate Type Ia X-ray images from Type CC.   
For future work on separating Type Ia from Type CC SNR images, machine learning methods are promising. 
For example, cluster analysis \citep{2022A&A...665A..26M} is an unsupervised machine learning algorithm that works on unlabelled data and partitions data into groups based on data similarity. This could distinguish SNR types purely based on the image properties and possibly reveal different distinctions than the bimodel Type Ia vs Type CC classification. 

{\large \noindent Acknowledgements}
This work supported by a grant from the Natural Sciences and Engineering Research Council of Canada.


\begin{thebibliography}{}

\bibitem[Arfken(1985)]{mathphys2}
Mathematical Methods for Physicists
G, Arfken
Published 1985 by Academic Press, Orlando
ISBN 0-12-059820-5

\bibitem[Buote \& Tsai(1995)]{1995ApJ...452..522B} Buote, D.~A. \& Tsai, J.~C.\ 1995, The Astrophysical Journal, 452, 522. doi:10.1086/176326

\bibitem[Chevalier(1982)]{1982ApJ...258..790C} Chevalier, R.~A.\ 1982, The Astrophysical Journal, 258, 790. doi:10.1086/160126

\bibitem[Cioffi et al.(1988)]{1988ApJ...334..252C} Cioffi, D.~F., McKee, C.~F., \& Bertschinger, E.\ 1988, The Astrophysical Journal, 334, 252. doi:10.1086/166834

\bibitem[Cox(2005)]{2005ARA&A..43..337C} Cox, D.~P.\ 2005, Annual Reviews of Astronomy and Astrophysics, 43, 337. doi:10.1146/annurev.astro.43.072103.150615

\bibitem[Dohm-Palmer \& Jones(1996)]{1996ApJ...471..279D} Dohm-Palmer, R.~C. \& Jones, T.~W.\ 1996, The Astrophysical Journal, 471, 279. doi:10.1086/177968

\bibitem[Dwarkadas(2007)]{2007ApJ...667..226D} Dwarkadas, V.~V.\ 2007, The Astrophysical Journal, 667, 226. doi:10.1086/520670

\bibitem[Ferreira \& de Jager(2008)]{2008A&A...478...17F} Ferreira, S.~E.~S. \& de Jager, O.~C.\ 2008, Astronomy and Astrophysics, 478, 17. doi:10.1051/0004-6361:20077824

\bibitem[Fukushima et al.(2020)]{2020ApJ...897...62F} Fukushima, K., Yamaguchi, H., Slane, P.~O., et al.\ 2020, \apj, 897, 62. doi:10.3847/1538-4357/ab94a6

\bibitem[Leahy et al.(2019)]{2019AJ....158..149L} Leahy, D., Wang, Y., Lawton, B., et al.\ 2019, The Astronomical Journal, 158, 149. doi:10.3847/1538-3881/ab3d2c

\bibitem[Lopez et al.(2009)]{2009ApJ...706L.106L} Lopez, L.~A., Ramirez-Ruiz, E., Badenes, C., et al.\ 2009, The Astrophysical Journal, 706, L106. doi:10.1088/0004-637X/706/1/L106

\bibitem[Lopez et al.(2011)]{2011ApJ...732..114L} Lopez, L.~A., Ramirez-Ruiz, E., Huppenkothen, D., et al.\ 2011, The Astrophysical Journal, 732, 114. doi:10.1088/0004-637X/732/2/114

\bibitem[Mahlke et al.(2022)]{2022A&A...665A..26M} Mahlke, M., Carry, B., \& Mattei, P.-A.\ 2022, \aap, Asteroid taxonomy from cluster analysis of spectrometry and albedo, 665, A26. doi:10.1051/0004-6361/202243587

\bibitem[Ranasinghe \& Leahy(2019)]{2019arXiv190911803R} Ranasinghe, S. \& Leahy, D.\ 2019, Journal of High Energy Physics, Gravitation, and Cosmology, 5, 907-919. doi:10.4236/jhepgc.2019.53046 

\bibitem[Riley, Hobson \& Bence(2006)]{mathphys1}
Mathematical Methods for Physics and Engineering
K.F. Riley, M.P. Hobson, S.J. Bence
Published March 13, 2006 by Cambridge University Press
ISBN 9780521679718 (ISBN10: 0521679710)

\bibitem[Truelove \& McKee(1999)]{1999ApJS..120..299T} Truelove, J.~K. \& McKee, C.~F.\ 1999, The Astrophysical Journal Supplement Series, 120, 299. doi:10.1086/313176

\end{thebibliography}
\end{document}